\documentstyle[psfig,aps,prl,amsfonts,amssymb]{revtex}

\begin{document}
\title{Mixed State Holonomies}
\author{Paul B. Slater}
\address{ISBER, University of
California, Santa Barbara, CA 93106-2150\\
e-mail: slater@itp.ucsb.edu,
FAX: (805) 893-7995}

\date{\today}

\draft
\maketitle
\vskip -0.1cm

\begin{abstract}
Sj\"oqvist, Pati, Ekert, Anandan, Ericsson, Oi and Vedral 
(Phys. Rev. Lett. 85, 2845 [2000]) have recently 
``provided a physical prescription based on interferometry for 
introducing the total phase of a mixed state undergoing unitary 
evolution, which has been an elusive concept in the past''.
They note that ``Uhlmann was probably the first to address the issue of
mixed state holonomy, but as a purely mathematical problem''.
We investigate 
possible relationships between these ``experimental'' and ``mathematical'' 
approaches, 
by examining various quantum-theoretic scenarios. 
We find that the two methodologies, in general, 
yield inequivalent outcomes.
\end{abstract}
\vspace{.1cm}



\vspace{.1cm}
\section{Introduction}
In the introductory paragraph of their recent 
interesting letter, Sj\"oqvist {\it et
al}  state that ``Uhlmann was probably the first to address
the issue of mixed state holonomy, but as a {\it purely mathematical}
(emphasis added)
problem. In contrast, here we provide a new formalism of the geometric
phase for mixed states in the {\it experimental} (emphasis added) 
 context of quantum
interferometry'' \cite{sjo1}. Left unaddressed there, however, 
 was the nature of any possible 
relationship between the two approaches. This question merits
 particular attention since Dittmann has ``shown that the 
connection form (gauge field) related to the generalization of the Berry
phase to mixed states proposed by Uhlmann satisfies the source-free
Yang-Mills equation $*D*D\omega=0$, where the Hodge operator is taken
with respect to the Bures metric on the space of finite-dimensional
nondegenerate density matrices'' \cite{ditt1}.
This fact, Dittmann and Uhlmann noted, 
``may be seen as an extension to mixed states of numerous examples
relating the original Berry phase to Dirac monopoles, and the Wilczek
and Zee phase, to instantons'' \cite[sec. 1]{dittuhl}.  In their work,
Sj\"oqvist {\it et al} do  
express the geometric phase in terms of an
{\it average} connection form \cite[eq. (16)]{sjo1} (for which no
Yang-Mills characterization has been given).
Zee  has reported  that the non-Abelian gauge structure associated
with nuclear quadrupole resonance  does {\it not} satisfy the source-free
Yang-Mills equation \cite{zee1} (cf. \cite{avron}). 
In \cite{slater}, in line with certain work of Figueroa-O'Farrill concerning 
calibrated geometries
\cite[sec. 6]{Fig}, we have begun 
 a study of the  
``Dittmann-Bures/Yang-Mills'' field over the {\it eight}-dimensional
convex set of {\it three}-level density matrices (making use of the
formulas for the Bures metric developed in \cite{slater4}, based on 
the parameterization in \cite{mark}) by 
finding certain self-dual
{\it four}-forms ($\varphi$) with respect to the Bures metric. The eigenspace
decomposition of the associated endomorphism $\hat{\varphi}$ 
was found to yield
four octets and three singlets as factors \cite{slater}.

The research presented below represents our effort to gain
a formal understanding of the relations between
the ``mathematical'' methodology of Uhlmann and 
the ``experimental'' approach of Sj\"oqvist {\it et al}. We hope that 
this will prove
useful, among other things, 
 in the devising of a physical scheme for implementing
the concepts of Uhlmann (cf. \cite[sec. 4]{sjo2}).

In the interferometric/experimental 
 case \cite{sjo1}, each pure state that diagonalizes
the initial density matrix ($\rho(0)$) is 
parallel transported separately from other distinct 
diagonalizing pure states,
so one simply weights or 
(incoherently) averages 
 the individual pure state 
(single-state) interference profiles by the corresponding 
eigenvalues of $\rho(0)$ to obtain 
the mixed state results . 
In the notation of \cite{sjo1}, the {\it total}
geometric phase is the argument
of $\sum_{k} w_{k} \nu_{k} e^{i \phi_{k}}$, and the visibility 
is its absolute
value, where $w_{k}, \nu_{k}, \phi_{k}$ are, respectively, 
the eigenvalue, visibility and geometric phase obtained for the $k$-th 
constituent pure state of $\rho(0)$ 
\cite[eqs. (10), (15)]{sjo1}.
The phase introduced by Sj\"oqvist {\it et al} ``fulfills two central
properties that make it a natural generalization of the pure 
state case:
(i) it gives rise to a linear shift of the interference oscillations
produced by a variable $U(1)$ phase and (ii) it reduces to the
Pancharatnam connection for pure states'' \cite{sjo1}.
The mixed state generalization of
Pancharatnam's connection --- which asserts that $\rho(t)$ and
$\rho(t + dt)$ are in phase if $\mbox{Tr}[\rho(t) U(t +dt) \dagger{U}(t)]$ 
is real and positive --- can be met only when \cite[eq. (12)]{sjo1}
\begin{equation}
\mbox{Tr} [\rho(t) \dot{U}(t) {U}^{\dagger}(t)]=0.
\end{equation}
This is the parallel transport condition 
of Sj\"oqvist {\it et al} for mixed states undergoing
unitary evolution.

In
the approach of Uhlmann \cite{uhl1}  
the parallel transport takes place
in a larger Hilbert space, defined by purification.
The geometric phase of Sj\"oqvist {\it et al} can also be understood using
purification \cite{sjo1}, though 
the unitary operator that acts on the ancilla 
or auxiliary state there is unconstrained and 
can in fact be anything, such as the identity operator. A 
``parallel transport of a density operator amounts to a parallel
transport of {\it any} (emphasis added) of it 
purifications \cite[p. 2848]{sjo1}.

In infinitesimal form, the parallelity condition of Uhlmann for a 
density matrix $\rho(t)$ reads \cite[eq. (19)]{sjo2},
\begin{equation} \label{w1}
\dot{W(t)^{\dagger}}  W(t)  = \mbox{hermitian},
\end{equation}
where $W(t) =\sqrt{\rho(t)} V(t)$. The unitary operator $V(t)$ 
can be interpreted
as acting on the ancilla in the purification of 
$\rho(t)$ 
and is determined by this parallelity
condition (\ref{w1}). For example, if the system, initially in the state
$\rho(0)$,  evolves unitarily under the evolution operator $U(t)$ one may
write $W(t) = U(t) \sqrt{\rho(0)} V(t)$. The 
Uhlmann parallelity condition, then,
 reads
\begin{equation} \label{w2}
\sqrt{\rho(0)} U^{\dagger}(t)  \dot{U}(t)  \sqrt{\rho(0)} =
V(t)  \dot{V}(t) ^{\dagger} \rho(0) + \rho(0) V(t)  \dot{V}(t)^{\dagger}.
\end{equation}
With given input $\rho(0)$ and $U(t)$ this 
equation needs to be solved for $V(t)$. 
The geometric phase $\gamma_{g}$ for any 
specific $T$ can then be computed using the mixed
state Pancharatnam connection \cite{sjo1},
\begin{equation} \label{w3}
\gamma_{g}(T) = \arg \mbox{Tr}
 (V^{\dagger} (T) \sqrt{\rho(0)} U^{\dagger}(T)
\sqrt{\rho(0)}).
\end{equation}

In the three sets of analyses reported below (sec. II), we will implement
certain formulas of Uhlmann \cite{uhl1,uhlparr}.
We compare the two methodologies under consideration in
the context of: (a)  unitary evolution over geodesic spherical 
{\it triangles}, the
vertices of which are associated with certain specific  
{\it mixed} two-level systems; (b) unitary evolution 
of such two-level systems over certain 
circular paths ($O(3)$-orbits); and (c) 
unitary evolution over such circular paths 
of $k$-level Gibbsian density matrices \cite{uhlparr,slater2}.
In case (a) we find an interesting correspondence between the two
approaches for mixed
states having Bloch vectors of length $\sqrt{{2 \over 3}}$ 
and in (c) that the results of the Uhlmann methodology
converge to those of Sj\"oqvist {\it et al} if 
 a certain variable (a), a function ($\mbox{sech}{\alpha \over 2}$)  
of the inverse temperature
parameter ($\alpha$), is driven to zero.
If instead of strictly mixed states, we consider the unitary evolution
of {\it pure} states, then both methodologies simply reproduce the
corresponding Berry phase for each of the three scenarios.
However, for general mixed states, the two methodologies typically 
yield inequivalent results.

\section{Analyses}

\subsection{Unitary evolution over geodesic spherical triangles}

For a qubit (a spin-${1 \over 2}$ particle), whose density matrix can be
written as,
\begin{equation} \label{fee}
\rho (0)= {1 \over 2} (1 + r \mathbf{\hat{r}} \cdot \mathbf{\sigma}),
\end{equation}
where $\mathbf{\hat{r}}$ is a 
unit vector, $r$ is the constant 
(length of the Bloch vector) for unitary evolution and $\mathbf{\sigma}$
is a vector of the three (non-identity) Pauli matrices,
Sj\"oqvist {\it et al} obtained a formula for the 
{\it geometric phase} \cite[eq. (25)]{sjo1},
\begin{equation} \label{one1}
 \gamma_{g}[\Gamma] = - \mbox{arctan} \Big( r \mbox{tan} {\Omega 
\over 2} \Big).
\end{equation}
Here $\Gamma$ is the unitary curve in parameter space 
traversed by the system 
and $\Omega$ the solid angle 
subtended by the natural extension of 
$\Gamma$ to the unit (Bloch) sphere. This 
formula reduces to $- {\Omega \over 2}$ for pure states ($r = 1$), as it, of
course,  
should.
The {\it visibility}  was expressed as
\cite[eq. (26)]{sjo1},
\begin{equation} \label{SJOVIS}
\nu = \eta \sqrt{\cos^{2}{\Omega \over 2} + r^2 \sin^{2}{\Omega \over 2}}.
\end{equation}
For the case of cyclic evolution, which will be our only area of 
concern here,
$\eta =1$.

We focus on the case of cyclic evolution for 
geodesic triangles, in particular, 
since Uhlmann has an enabling formula ``relevant 
for the intensities'',
\cite[eq. (24)]{uhl1}
\begin{equation} \label{Armin1}
\sqrt{a_{12} a_{23} a_{31}} \mbox{Tr} \rho (0) U = 
r^4 a_{321} +r^3 (1-r) {a_{12} + a_{23} +a_{31} + 2 a_{321} -3 
\over 2} +
\end{equation}
\begin{displaymath}
+ r^2 (1-r)^2 {a_{12} + a_{23} + a_{31} + 6 \over 2} + 4 r (1-r)^3 + (1-r)^4.
\end{displaymath}
Here
\begin{equation} \label{Armin2}
a_{12}:=  \langle 1|2 \rangle  \langle 2|1 \rangle ,\quad  a_{321}=  \langle 3|2 \rangle  \langle 2|1 \rangle  \langle 1|3 \rangle , \ldots,
\end{equation}
where $|j  \rangle   \langle  j|$ gives the projection operator 
$P_{j}$ associated with the $j$-th
density matrix ($\tilde{\rho}_{j}$) in the geodesic triangle, that is, 
\begin{equation}
P_{j} = {2 \tilde{\rho}_{j} - (1-r) \underline{1}  \over 2 r}.
\end{equation}
 Uhlmann also apparently has an explicit  formula for
geodesic {\it quadrangles},  but only the leading terms of it have been 
published  \cite[eq. (25)]{uhl1}, that is,
\begin{equation} \label{Armin3}
\sqrt{a_{12} a_{23} a_{34} a_{41}} \mbox{Tr} \rho (0) U = r^5 a_{4321} +
\end{equation}
\begin{displaymath}
+ r^4 (1-r) {6 a_{4321} -a_{432} - a_{421} + 3 a_{431} +
3 a_{321} + a_{43} + a_{21} + a_{41} + a_{23} + a_{42}
- 3 a_{13} + 2 \over 2} + \ldots.
\end{displaymath}

Without loss of generality, let us set one of the three vertices 
of the geodesic spherical triangle 
swept out by the unit vector $\mathbf{\hat{r}}$  (\ref{fee}) in the 
unitary evolution to be
(0,0,1). We use the standard spherical coordinate
form for a point on the sphere, that is 
$(\sin{\theta} \cos{\phi},
\sin{\theta} \sin{\phi},\cos{\theta})$, and take the other
two vertices to be parameterized by the 
angular pairs $(\theta_{1},\phi_{1})$ 
and $(\theta_{2},\phi_{2})$, so  the 
preassigned vertex (0,0,1) corresponds simply to
$\theta_{1}  = 0$.
Then, using the notation
$\bar{\phi} = \phi_{1} - \phi_{2}$, we have for the solid angle 
\cite{Eriksson},
\begin{equation} \label{solid}
\Omega = - 2 \mbox{arccot} \Big(\cot{\bar{\phi}} + 
(\cot{\theta_{1}} + \csc{\theta_{1})} 
(\cot{\theta_{2}} +\csc{\theta_{2}}) \csc{\bar{\phi}} \Big).
\end{equation}
(This formula, as well as all the succeeding ones here, 
depend implicitly or explicitly
upon $\bar{\phi}$ and not on 
$\phi_{1}$ and $\phi_{2}$ individually.)
Let us also denote
\begin{equation}
\mu= (1+ \cos{\theta_{1}}) (1+\cos{\theta_{2}}) + \cos{\bar{\phi}}
\sin{\theta_{1}} \sin{\theta_{2}}.
\end{equation}
The parameter $\mu$ is confined to the range $[-{1 \over 2},4]$, 
for $0 \leq \theta_{1},\theta_{2} \leq \pi, \quad 0 \leq \phi_{1},\phi_{2} 
< 2 \pi$. It is equal to its minimum 
-${1 \over 2}$ at $\theta_{1} = \theta_{2} =
{2 \pi \over 3}$ and $\bar{\phi} = \pi$; equal to its maximum 4
for $\theta_{1} = \theta_{2} = 0$
and equal to zero if {\it either} $\theta_{i} = \pi$ 
or $\theta_{2} = \pi$.

Making use of (\ref{Armin1}) and 
(\ref{Armin2}), we have for the ``Uhlmann geometric phase'' 
for the  general geodesic triangle specified that 
\begin{equation} \label{gp}
\tilde{\gamma}_{g} = \mbox{arg}(\mbox{Tr} \rho_{1} U) = \mbox{arctan} \Big(
{\alpha \over \beta }
\Big),
\end{equation}
where
\begin{equation}
\alpha= r^3 \sin{\bar{\phi}} \sin{\theta_{1}} \sin{\theta_{2}},
\qquad \beta= 4 +(\mu -10) r^2 + 6 r^4.
\end{equation}
We found that
\begin{equation}
  \tan{{\Omega \over 2}} = - {\alpha \over \mu r^3},
\end{equation}
so one can write  (\ref{one1})
\begin{equation}
\gamma_{g} =- \mbox{arctan} \Big( r \tan{{\Omega \over 2}} \Big) = 
\mbox{arctan} \Big( {\alpha \over \mu r^2} \Big),
\end{equation}
as well as
\begin{equation}
\tilde{\gamma}_{g} = \mbox{arctan} \Big( {\alpha \over \beta} \Big) = 
\mbox{arctan} \Big( {- \mu r^3 \tan{{\Omega \over 2}} 
\over \beta} \Big).
\end{equation}

Let us note the {\it exact} relations between the two forms 
(``mathematical'' and ``experimental'') of geometric
phase,
\begin{equation} \label{ratio}
{\mbox{tan} \Big( \gamma_{g} \Big)  \over 
\mbox{tan} \Big( \tilde{\gamma}_{g} \Big)} = 
1 + {4 -10 r^2 +6 r^4 \over 
\mu r^2} = 1 +{\beta - \mu r^2 \over \mu r^2} = {\beta \over \mu r^2},
\end{equation}
which are unity for both $r=1$ and $r = \sqrt{{2 \over 3}}$.
(Anandan {\it et al} \cite{andy} assert that ``a fully gauge-invariant
description of the geometric phase {\it requires} that it be defined
modulo $2 \pi$''. Of course, $\tan{\psi} =\tan{(\psi+\pi)}$.)
For  $r= 1 - \triangle r $, 
\begin{equation}
{\tan{\gamma_{g}} \over \tan{\tilde{\gamma}_{g}}} 
\approx \Big( 1 - {4 \triangle r 
\over \mu} \Big).
\end{equation}
For $r = \sqrt{{2 \over3}} +\triangle r$,
\begin{equation}
{\tan{\gamma_{g}} \over \tan{\tilde{\gamma}_{g}}} \approx
\Big( 1 - {2 \sqrt{6} \triangle r \over \mu} \Big).
\end{equation}
For $r= \triangle r$ (that is, a neighborhood of the fully mixed 
[classical] state, 
$r=0$),
\begin{equation} \label{finalone}
{\gamma_{g} \over \tilde{\gamma}_{g}} \approx { \tan{\gamma_{g}} \over
\tan{\tilde{\gamma}_{g}}} \approx 1 +{ 4 -10 \triangle^2 r \over
\mu \triangle^2 r }.
\end{equation}

The ``Uhlmann visibility'' is 
\begin{equation}
\tilde{\nu} = |\mbox{Tr} \rho_{1} U| =\sqrt{{\alpha^2 +\beta^2 \over (\alpha
/  r^3)^2 + \mu^2}},
\end{equation}
which is precisely unity for $r=1$, as is also the case 
for $\nu$, given by  (\ref{SJOVIS}).
Now,
\begin{equation} \label{ratio2}
{\nu \over \tilde{\nu}} = {1 \over r^2} \sqrt{{\alpha^2 +  \mu^2 r^4 \over 
\alpha^2 + \beta^2}}.
\end{equation}
This ratio is always unity for $r=1$ and always ${3 \over 2}$ for
$r=\sqrt{{2 \over 3}}$. (Recall, on the other hand,
 that the ratio (\ref{ratio}) of the tangents of the
two geometric phases is {\it unity} for $r =\sqrt{{2 \over 3}}$.)
For $r =1 -\triangle r$,
\begin{equation}
{\nu \over \tilde{\nu}} \approx 1 + \triangle r \Big( 2 + {4 \mu \over
(\alpha/r^3)^2 + \mu^2}  \Big).
\end{equation}
For $r = \sqrt{{2 \over 3}} + \triangle r$,
\begin{equation}
{\nu \over \tilde{\nu}} \approx {3 \over 2} \Big( 1 + \sqrt{6} \triangle r
({6 \mu \over 2 (\alpha / r^3)^2 + 3 \mu^2} -1) \Big).
\end{equation}
For $r = \triangle r$,
\begin{equation}
{\nu \over \tilde{\nu}} \approx {\mu \over 4} + 
\triangle^2 r \Big( {(\alpha / r^3)^2 \over 8 \mu} - {1 \over 16} 
(\mu -10) \mu \Big).
\end{equation} 
In the limit $r \rightarrow 0$,
\begin{equation}
{\nu \over \tilde{\nu}} = {\mu \over 4}.
\end{equation}
\subsection{Unitary evolution over certain circular paths}

\begin{figure}
\centerline{\psfig{figure=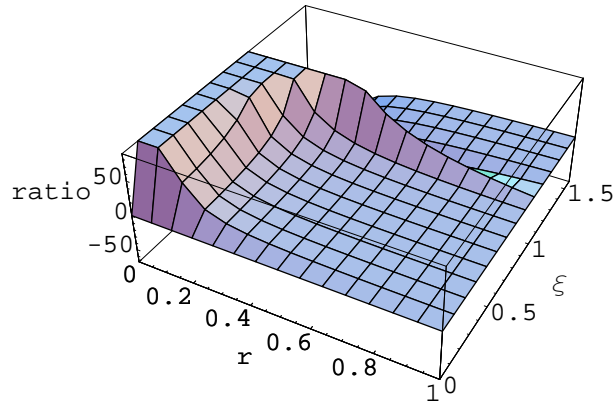}}
\caption{Ratio (\ref{par1}) of tangents of geometric phases for
unitary evolution over circular paths}
\label{lb}
\end{figure}

\begin{figure}
\centerline{\psfig{figure=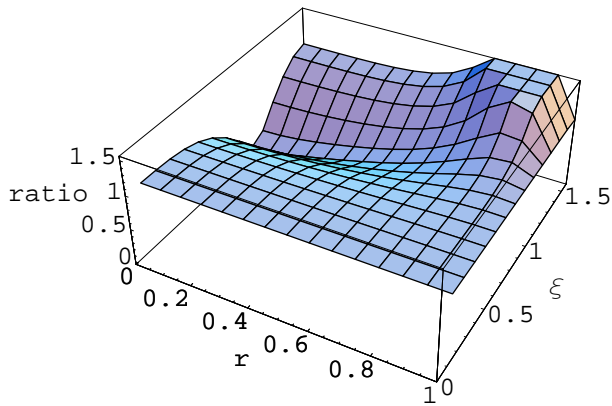}}
\caption{Ratio (\ref{par2}) of visibilities for unitary
evolution over circular paths}
\label{lc}
\end{figure}

Now let us similarly study the unitary evolution of spin-${1 \over 2}$
systems over {\it circular} ($O(3)$-orbits) --- as opposed 
to geodesic triangular --- paths.
We utilize the analyses of Uhlmann in \cite{uhlparr}.
Again, we start with an initial density matrix $\rho(0)$, 
the Bloch vector of
which is proportional to (0,0,1), while the axis of rotation about
which the state evolves is $\overrightarrow{n}
 =(0,\sin{\xi},\cos{\xi})$.
The resulting curve of density matrices is given by
\begin{equation}
\phi \rightarrow U(\phi) \rho(0) U(-\phi), \qquad U(\phi) =
e^{- i \phi (\sin{\xi} J_{y} + \cos{\xi} J_{z})},
\end{equation}
and the associated parallel lift of this curve with initial value
${\rho (0)}^{1/2}$ by
\begin{equation}
\phi \rightarrow U(\phi) {\rho(0)}^{1/2} V(\phi), \qquad V(\phi) =
e^{i \phi (\sin{\xi} \hat{J}_{y} + \cos{\xi} J_{z})},
\end{equation}
where
\begin{equation}
2 \rho(0)^{1/2} J_{y} \rho(0)^{1/2} = \rho(0) \hat{J}_{y}
+\hat{J}_{y} \rho(0),
\end{equation}
and the $J$'s are, of course, generators of an irreducible representation
of $SU(2)$.
We have the relation
\begin{equation}
\xi =  \cos^{-1}{\Big( - {\Omega \over 2 \pi}\Big) } .
\end{equation}
The associated holonomy invariant for a complete rotation
about $\overrightarrow{n}$ is 
\begin{equation}
 \rho(0)^{1/2} U(2 \pi) V(2 \pi) \rho(0)^{1/2}.
\end{equation}
(In \cite[sec. V]{slater2}, we have studied certain ``higher-order'' holonomy
invariants.)
We then have the result (using the ``$\quad \breve{} \quad $'' notation 
now, rather 
than ``$\quad \tilde{} \quad$'' as before, to refer
to results obtained based on the unitary circular 
evolution analysis of Uhlmann \cite{uhlparr})
\begin{equation} \label{par1}
{\tan{\gamma_{g}} \over \tan{\breve{\gamma}_{g}}} = 
 \chi  \sec{\xi} \coth{\pi \chi} \tan{(\pi \cos{\xi})} = 
{\pi \chi  \tan{\kappa} \coth{\pi \chi} 
\over \kappa },
\end{equation}
where
\begin{equation}
\chi = \sqrt{{r^2 - 2 -r^2 \cos{2 \xi} \over 2}},
\end{equation}
and 
\begin{equation}
\kappa = {\pi \sqrt{-1 + r^2 -\chi^2} \over r} = \pi \cos{\xi} = 
-{\Omega \over 2}.
\end{equation}
We plot this ratio (\ref{par1}) in Fig.~\ref{lb}.
Now for $\triangle \xi$,
\begin{equation}
{\tan{\gamma_{g}} \over \tan{\breve{\gamma}_{g}}}
\approx {4 + 3 ( 1-r^2) (\triangle \xi)^2 \over 4 r^2}.
\end{equation}
For pure states ($r=1$), the ratio (\ref{par1}) reduces to 1.
  The ratio of the visibilities,
\begin{equation} \label{par2}
{\nu \over \breve{\nu}} = 
{\chi \sqrt{2} \sqrt{ \cos^{2}{\kappa} + r^2 \sin^{2}{\kappa}} \over
\sqrt{ 1- r^2 + 2 \chi^2 +(r^2 -1) \cosh{ 2 \pi \chi}}} =
{\chi \sqrt{2}  \sqrt{\cos^{2}{ \kappa} +
(1+\chi^2) \csc^{2}{\xi} \sin^{2}{\kappa }} \over
\sqrt{1 + 2 \chi^{2} - \cosh{2 \pi \chi} + 2
(1 +\chi^2) \csc^{2}{\xi} \sinh^{2}{\pi \chi}}},
\end{equation}
 is given in Fig.~\ref{lc}.
Both the ratios (\ref{par1}) and (\ref{par2}) are unity for
pure states ($r=1$).
\subsection{Unitary evolution of Gibbsian $k$-level 
density matrices} \label{ig}

\begin{figure} \label{y3}
\centerline{\psfig{figure=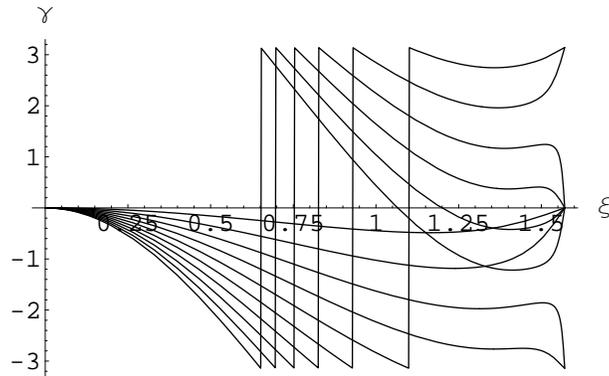}}
\caption{Geometric phases, based on the Uhlmann methodology, for the
$k$-level Gibbsian density matrices, holding the inverse temperature
parameter $\alpha$ fixed at 2. Below $\xi \approx .65$, the curve for
$k=2$ dominates that for $k=3$, which dominates that for $k=4,\ldots$}
\label{p1}
\end{figure}

\begin{figure}
\centerline{\psfig{figure=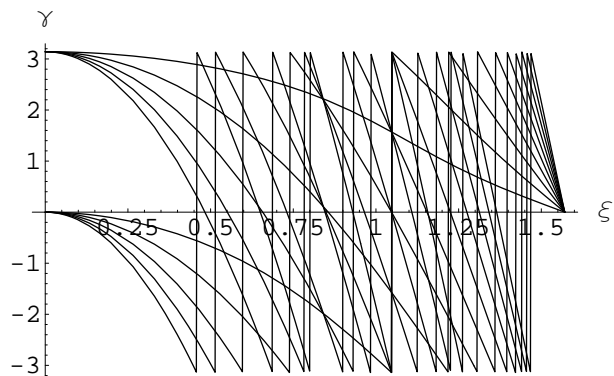}}
\caption{Geometric phases, based on the Sj\"oqvist {\it et al}
methodology, for the $k$-level Gibbsian density matrices, holding the
inverse temperature parameter $\alpha$ fixed at 2. Below 
$\xi \approx .45$, 
the upper (even $k$) group of curves in decreasing order of
dominance correspond to $k=2,4,6,8,10$
and the lower (odd $k$) group to $k=3,5,7,9,11$}
\label{p2}
\end{figure}

In the final part of the 
paper \cite{uhlparr}, Uhlmann also considered the analogous unitary
circular  evolution ($O(3)$-orbits) 
of $k$-level Gibbsian states of the form
\begin{equation}
\rho = {e^{\alpha \overrightarrow{n} \overrightarrow{J}} \over
\mbox{trace} \quad e^{\alpha \overrightarrow{n} \overrightarrow{J}}},
\end{equation}
where the vector $\overrightarrow{J}$ of angular momentum operators lives
in $k$-dimensions.
(For $k=2$, the model is equivalent to that studied in the previous
section.)
In \cite{slater2} we have conducted a detailed analysis of this model in
comparison with the results that would be 
obtained implementing the
methodology of Sj\"oqvist {\it et al}. One unanticipated outcome was that
the Sj\"oqvist {\it et al} model appeared sensitive to the bosonic
(odd $k$) or fermionic (even $k$) nature of the $k$-level systems.
As an example of this phenomenon, we exhibit two additional plots
(Fig.~\ref{p1} and \ref{p2})
based on analyses for $k=2,\ldots,11$. 
In each of these plots, the inverse temperature
parameter $\alpha$ is held fixed at 2. In Fig.~\ref{p1}, based on
the Uhlmann methodology, below 
$\xi \approx .65$ the ten curves are all monotonically decreasing. 
The order of dominance among themselves of these curves is the natural order
$k=2,\ldots,11$, with the curve for $k=2$ being the dominant one. 
On the other hand, using the procedure of
Sj\"oqvist {\it et al}, the ten 
corresponding curves are separated into two 
clusters of five (below 
$\xi \approx .44$) (Fig.~\ref{p1}). 
In the first cluster (having the higher values of $\gamma$) 
in decreasing order of dominance the five curves correspond to 
$k=2,4,6,8,10$, while in the second group (having lesser 
values) 
the order is $k=3,5,7,9,11$. (For other related Figures, 
see \cite[sec. VIII]{slater2}.)
Also, we have found \cite{slater2} that the (first-order) holonomy 
invariant in the Uhlmann approach \cite[eqs. (33), (43), (47)]{uhlparr}, 
that is the trace of
\begin{equation} \label{q1}
 W(2 \pi) W(0)^{\dagger} = 
(-1)^{2 j} \rho(0)^{1/2} e^{2 \pi i (\cos{\theta} J_{z}
+a \sin{\theta} J_{y})} \rho(0)^{1/2},
\end{equation}
 reduces to $(-1)^{k+1}$ times
the holonomy invariant \cite[eq. (15)]{sjo1} 
in the Sj\"oqvist {\it et al} analysis if we simply
set the parameter $a = \mbox{sech}{\alpha \over 2}$ in (\ref{q1}) 
to zero, that is the influence of the angular momentum
operator $J_{y}$ is nullified. Relatedly, if $\rho(0)$ were 
the density matrix of a {\it pure} state, 
then the term $a \sin{\theta} J_{y}$ in the invariant (\ref{q1}) 
would drop out, with the trace of the resultant expression being
the corresponding Berry phase \cite[eq. (43)]{uhlparr}.
This fact was employed in \cite{slater2} to compute the Sj\"oqvist
{\it et al} mixed state holonomy for the $k$-level Gibbsian density 
matrices, by taking the 
average  (using the eigenvalues of $\rho(0)$ as weights) of these 
Berry phases for the pure states that diagonalize the initial density
matrix \cite[eqs. (10), (15)]{sjo1}.

\section{Concluding Remarks}

The analyses presented above clearly demonstrate a general
inequivalence between the ``experimental'' approach to mixed state holonomy
of Sj\"oqvist {\it et al} \cite{sjo1} 
and the ``purely mathematical'' scheme of Uhlmann
\cite{uhl1,uhlparr}. In particular, in the third of our analyses
(sec.~\ref{ig}), concerned
with $O(3)$-orbits of $k$-level Gibbsian 
density matrices, it has emerged that
the analyses of Sj\"oqvist {\it et al} (Fig.~\ref{p2}) 
are sensitive to the parity 
of $k$, while those of Uhlmann (Fig.~\ref{p1}) are not \cite{slater2}.
 (For an application of \cite{sjo1} 
to the detection 
of quantum entanglement, see \cite{horeke}. For a further discussion of
the results in \cite{sjo1}, see \cite[secs. 2, 4]{sjo2}.)

A problem that remains to be fully addressed is the conceptualization of
a physical apparatus 
that would yield
 the geometric phases and visibilities given  by the analyses
of Uhlmann \cite{uhl1,uhlparr}. 
In their study \cite{sjo1}, Sj\"oqvist {\it et al} did describe the use
of a Mach-Zehnder interferometer to ``test the geometric phase for
mixed states and provide the notion of phase and parallel transport
for mixed states in a straightforward way''. However, they did not 
specify such a physical scheme for their conceptual alternative
based on a purification procedure (cf. \cite[sec. 4]{sjo2}).

Let us also note that the connection 
used for parallel transport in the Uhlmann (Bures) case 
is but one of an (uncountable)
infinite set of 
connections 
(``defining reasonable parallel transports along curves of
density operators''), which correspond in a one-to-one fashion,
with the {\it monotone} metrics on the $k$-level density matrices, and are
all compatible with the purification procedure 
\cite{dittuhl,petzsudar,slaterLLMMPP}. In this 
generalized framework, the 
Uhlmann (canonical) connection corresponds
to the well-studied Bures or minimal monotone metric \cite{ditt1,ditt}. 
A yet apparently formally unsettled question is 
whether or not the Sj\"oqvist {\it et al} connection
\cite[eqs. (16)-(18)]{sjo1} itself corresponds
 to some such monotone metric, thus falling within this  generalized
scheme of Dittmann and Uhlmann  of connections (and metrics) 
respecting standard purification \cite{dittuhl}. 
From the results presented above, it is clear
 that
it does {\it not} correspond to the Bures metric.
(In the Uhlmann methodology, ``proper liftings'' $W(t)$ are those that
minimize the length functional $\int \mbox{Tr} \mbox [d \dot{W}^{*}dW]$ 
of the curve of density operators undergoing unitary evolution. 
For a proper lifting this length becomes equal to the ``Bures length''.
One particularly interesting property of the Bures metric is that its
volume element can be employed to determine analytically {\it exact}
 probabilities
that a pair of quantum bits is [{\it a priori}] classically 
correlated \cite{slater3}.)

\acknowledgments

I would like to express appreciation to the Institute for Theoretical
Physics for computational support in this research and to E.
Sj\"oqvist for his helpful correspondence.

\end{document}